\documentclass{ecai}

\usepackage{times}
\usepackage{soul}
\usepackage{url}
\usepackage[hidelinks]{hyperref}
\usepackage[utf8]{inputenc}
\usepackage[small]{caption}
\usepackage{algorithm}
\usepackage[numbers, sort, comma, square]{natbib}
\usepackage{graphicx}
\usepackage{amsmath}
\usepackage{amssymb} 
\usepackage{amsthm}
\usepackage{balance}
\usepackage{subfigure}
\usepackage{float}
\usepackage{dsfont}
\usepackage{algpseudocode}
\usepackage{booktabs}
\usepackage{latexsym}

\def\IR{\mathds{R}}

\newcommand{\A}{\mathcal{A}}

\renewcommand{\H}{\mathcal{H}}
\newcommand{\M}{\mathcal{M}}

\newcommand{\X}{\mathcal{X}}

\newcommand{\Z}{\mathcal{Z}}

\newcommand{\eps}{\varepsilon}

\renewcommand{\P}{\mathbf{P}}
\renewcommand{\O}{\mathbf{O}}
\renewcommand{\vec}[1]{{\boldsymbol#1}}

\newcommand{\EE}[2][]{\mathbb{E}_{#1}\left[#2\right]} 
\newcommand{\PP}[2][]{\mathbb{P}_{#1}\left[#2\right]} 

\providecommand{\set}[1]{\ensuremath{\left\{#1\right\}}}

\newcommand{\bel}{\mathrm{Bel}}
\newcommand{\KL}{\mathrm{KL}}

\theoremstyle{plain}
\newtheorem{proposition}{Proposition}
\newtheorem{lemma}[proposition]{Lemma}

\DeclareMathOperator{\argmax}{\rm argmax}

\begin{document}

\begin{frontmatter}

\title{Making Friends in the Dark:\\
Ad Hoc Teamwork Under Partial Observability}

\author[A]{\fnms{João G.}~\snm{Ribeiro}\thanks{Corresponding Author. Email: joao.g.ribeiro@tecnico.ulisboa.pt}}
\author[A]{\fnms{Cassandro}~\snm{Martinho}}
\author[A, B]{\fnms{Alberto}~\snm{Sardinha}}
\author[A]{\fnms{Francisco S.}~\snm{Melo}}

\address[A]{INESC-ID \& Instituto Superior Técnico, University of Lisbon}
\address[B]{Pontifícia Universidade Católica do Rio de Janeiro}

\begin{abstract}
This paper introduces a formal definition of the setting of ad hoc teamwork under partial observability and proposes a first-principled model-based approach which relies only on prior knowledge and partial observations of the environment in order to perform ad hoc teamwork. We make three distinct assumptions that set it apart previous works, namely: i) the state of the environment is always partially observable, ii) the actions of the teammates are always unavailable to the ad hoc agent and iii) the ad hoc agent has no access to a reward signal which could be used to learn the task from scratch. Our results in 70 POMDPs from 11 domains show that our approach is not only effective in assisting unknown teammates in solving unknown tasks but is also robust in scaling to more challenging problems. Supplementary material is available at \textit{\url{https://github.com/jmribeiro/adhoc-teamwork-under-partial-observability}}
\end{abstract}

\end{frontmatter}

\section{Introduction}
One of many aspects of human intelligence is the ability to cooperate with strangers when unexpected events, such as emergencies, occur. Introduced by Stone et al. \citep{stone10aaai}, the setting of ad hoc teamwork, studies how autonomous agents can efficiently handle this same scenario. Instead of being evaluated according to their ability to learn and/or perform a given task, agents are evaluated according to how efficient they are able to assist unknown teammates in performing unknown tasks (without being able to pre-coordinate or communicate beforehand).

Even though the setting of ad hoc teamwork has recently been under great attention by the multi-agent systems community \citep{mirsky2022survey}, current approaches \citep{barrett17aij} rely on assumptions unfeasible to attain in real-world scenarios, such as full observability of the state of the environment \citep{barrett17aij, ribeiro2021helping} and being able to observe the actions of the teammates \citep{barrett17aij}. In a more recent line of work, Gu et al. \citep{gu2021online} showcase that by learning via trial-and-error how to perform tasks on a fully observable environment (where both the states and actions of the teammates are fully observable), an ad hoc agent is then able to transfer this knowledge to a partial observable environment where only a single observation is available. 

In this paper, we follow both Ribeiro et al. \citep{ribeiro2021helping} and Gu et al. \citep{gu2021online}, and formalize the problem of ad hoc teamwork under partial observability, relying on the novel assumptions that an agent never has access to the full state of the environment nor the actions of its teammates. We propose a novel, first-principled Bayesian online inference algorithm for the setting of ad hoc teamwork, named ATPO (\textbf{A}d hoc \textbf{T}eamwork for \textbf{P}artially \textbf{O}bservable environments), which relies on a library of possible models (in our case, Partially Observable Markov Decision Processes (POMDPs)). To the best of our knowledge, ATPO represents the first model-based approach for ad hoc teamwork capable of assisting unknown teams in completing unknown tasks, relying only on partial observations of the environment. These assumptions also make ATPO suited to address tasks involving {\em ad hoc robotic agents}, accommodating the natural perceptual limitations of robotic platforms \citep{genter2017jaamas} and allowing for the interaction, not only with other robots but also humans as teammates.

We evaluate ATPO by conducting an empirical evaluation, resorting to a total of 70 POMDPs from 11 benchmark multiagent problems. Our results demonstrate the effectiveness of ATPO in performing ad hoc teamwork under partial observability by not only efficiently assisting unknown teammates, but also identifying, from partial observations, which teammates it is interacting with and which tasks they are currently performing.

To summarize, our contributions are threefold:

\begin{itemize}
    \item We contribute to the formalization of ad hoc teamwork under partial observability;
    \item We propose ATPO, a novel approach for ad hoc teamwork under partial observability that is capable of inferring the underlying target task and teammate's behaviour from the agent's history of observations;
    \item We illustrate the applicability of our approach in a set of eleven distinct domains.
\end{itemize}

\section{Related Work}

The setting of ad hoc teamwork was first introduced by Stone et al. \cite{stone10aaai}. In ad hoc teamwork, an agent (named the {\em ad hoc agent}) has the goal of joining an already formed team and assisting it on-the-fly. The ad hoc agent is not assumed to be able to pre-coordinate or pre-communicate with the team.

Earlier work in the setting of ad hoc teamwork assumed that the ad hoc agent knew, beforehand, the teams' behavior and the task they were performing. Given this information, the only remaining goal for the ad hoc agent is to understand its role and act accordingly. Stone and Kraus \citep{Stone2010ToTeams} showcase an example of this scenario, introducing a finite-horizon decision problem where an ad hoc agent uses a planner in order to compute its own actions, assisting a single teammate in performing a single task. The authors are then followed by Barrett and Stone \citep{Barrett2011AdRewards}, who extend the approach to discounted infinite horizon problems. Later on, Agmon and Stone \citep{agmon12aamas} approach the same scenario, but now considering $N$ teammates instead of one. All these approaches, however, by assuming the ad hoc agents knows the teammates' behavior and the task they are performing beforehand, reduce ad hoc teamwork to a planning problem (where the only challenge is to figure out the optimal role for the ad hoc agent).

It is later on that Chakraborty and Stone \citep{chakraborty13aamas} drop the assumption that the teammates' behavior is known a priori. In their work, the authors assume that the teammates follow an unknown Markov policy that the ad hoc agent must identify to plan its actions. The authors, however, still assume that the task being performed is known to the ad hoc agent. From this perspective, ad hoc teamwork is closely related to {\em learning in games}, a problem for which an extensive literature already exists \citep{Fudenberg2000TheGames}.

More recently, Melo and Sardinha \citep{melo16jaamas} formalize the setting of ad hoc teamwork by breaking it down into three sub-problems: (i) task identification (the ad hoc agent has to identify the task the teammates are performing), (ii) teammate identification (the ad hoc agent has to identify the behavior of its teammates) and (iii) planning/execution (after identifying the task and teammate behavior, the ad hoc agent has to plan its best response accordingly). The authors are then followed by Barrett et al. \citep{barrett17aij}, who follow this formalization and introduce the PLASTIC framework \citep{barrett17aij, barrett15aaai}. The PLASTIC framework is an approach which, relying on a set of previous interactions, is able to identify the current team (and task being performed) by performing Bayesian updates. This is done by first assigning each possible model a belief prior, and, using as evidence the observed actions of the teammates for the current state, updating the beliefs in each timestep until. Using the most likely model, a pre-trained policy for the ad hoc agent can then be used to compute the best response.

Expanding upon Barrett et al. \cite{barrett17aij}, Ribeiro et al. \citep{ribeiro2021helping} later on tackle these limitations by dropping the assumption that the actions of the teammates are visible to the ad hoc agent. In their work, the authors also rely on a set of prior interactions, however this time modeled as Markov Decision Processes. Gu et al. \citep{gu2021online} afterwards introduce an approach relying on reinforcement learning techniques to allow an agent to first learn how to perform tasks from scratch in a fully observable environment (in which the agent is also able to observe the actions of its teammates) and is then deployed into a partial observable scenario, where it has to assist unknown teammates. In this paper, we expand upon Ribeiro et al. \citep{ribeiro2021helping} and Gu et al. \citep{gu2021online} by relying on the assumptions that the state of the environment is never observable and the actions of the teammates never visible to the ad hoc agent. Furthermore, we consider a setting of closed ad hoc teamwork, where teams are fixed and teammates do not dynamically join and leave. We formalize ad hoc teamwork under partial observability and propose a novel algorithm (ATPO) which assumes the environment is partially observable and the ad hoc agent has no access to the actions of its teammates. Similar to Ribeiro et al. \citep{ribeiro2021helping}, ATPO relies on a library of prior interactions, this time modeled as Partially Observable Markov Decision Processes (POMDPs). The beliefs over each possible model are then updated using Bayesian updates, using as evidence the partial observations. ATPO then uses the beliefs to compute the best response for the ad hoc agent, by relying on the solutions for the POMDPs, obtained using an arbitrary POMDP solver (in our case, Perseus \citep{spaan2005perseus}).

\subsection{Similar Lines of Research}

Four similar lines of work, parallel to ad hoc teamwork, are the assistance framework \citep{fern2007decision}, the problem of zero-shot coordination \citep{hu2020other}, the IPOMDP framework \citep{gmytrasiewicz2005framework} and the problem of few-shot coordination \citep{fosong2022few}. The {\em assistance} framework \citep{fern2007decision} models a scenario where an agent must help a teammate in solving a given sequential task under uncertainty. The goal, however, is for an agent to assist unknown teammates in performing known tasks. Even though the authors do not refer to the problem as ad hoc teamwork, one could argue that the identification of unknown teammates with known tasks falls under the scope of ad hoc teamwork. In their work, however, Fern et al. \cite{fern2007decision} also consider that the teammate's actions to be accessible to the assistant. The problem of {\em zero-shot coordination} \citep{hu2020other}, unlike ad hoc teamwork, studies how independently trained agents may interact with one another on first-attempt \citep{lupu2021trajectory,treutlein2021new,bullard2021quasi}. This can be seen as a flavor of ad hoc teamwork where there is only one teammate which the ad hoc agent knows to have computed an optimal solution to the task being performed. It also assumes the ad hoc agent knows the task being performed, reducing it only to a problem of {\em teammate identification}. The IPOMDP framework \citep{gmytrasiewicz2005framework}, considers how an agent can augment the state space of a POMDP taking into account all possible, unknown, teammates. Not only does this approach grow exponentially with the number of possible teammates, but, similar to the framework of assistance, it also assumes that all teammates perform the same task (modeled by a single reward function). In their work, however, Gmytrasiewicz and Doshi \citep{gmytrasiewicz2005framework} do not assume that the teammates' actions are visible to the agent, therefore sharing a common assumption with our work. Unlike Gmytrasiewicz and Doshi \citep{gmytrasiewicz2005framework}, our approach grows linearly with the number of possible teammates and tasks (instead of exponentially, like with IPOMDPs). Finally, the problem of few-shot teamwork (FST) considers a scenario where teams of agents are tasked with adapting to one another, similar to ad hoc teamwork. FST, however, unlike ad hoc teamwork, allows for an adaptation phase where teams can interact with one another before the final evaluation takes place, while ad hoc teamwork, requires agents to adapt on-the-fly without any pre-coordination or pre-communication.

\section{Background}%
\label{Sec:Background}

A {\em Markov decision problem} \citep{puterman05}, or MDP, is denoted as a tuple $(\X,\A,\set{\P_a,a\in\A},r,\gamma)$, where $\X$ is the state space, $\A$ is the action space, $\P_a$ is a transition probability matrix, where $\P_a(x'\mid x)$ is the probability of moving from state $x$ to $x'$ given action $a\in\A$, $r$ is the expected reward function, and $\gamma\in[0,1]$ is a discount factor.  

A {\em policy} $\pi$ maps states to distributions over actions. We write $\pi(a\mid x)$ to denote the probability of selecting action $a$ in state $x$ according to policy $\pi$. Solving an MDP consists of determining a policy $\pi$ to maximize the value
\begin{equation}
v^\pi(x)\triangleq\EE[A_t\sim\pi(X_t)]{\sum_{t=0}^\infty\gamma^tR_t\mid X_0=x},
\end{equation}
for any initial state $x\in\X$. In the above expression, $X_t$, $A_t$ and $R_t$ denote the (random) state, action and reward at time step $t$. The function $v^\pi:\X\to\IR$ is called a {\em value function}, and a policy $\pi^*$ is {\em optimal} if, given any policy $\pi$, $v^{\pi^*}(x)\geq v^\pi(x)$, for all $x\in\X$. The value function associated with an optimal policy is denoted as $v^*$ and can be computed using, for example, dynamic programming. An optimal policy, $\pi^*$, is such that $\pi^*(a\mid x)>0$ only if $a\in\argmax q^*(x,\cdot)$, where
\begin{equation*}
q^*(x,a)=r(x,a)+\gamma\sum_{x'\in\X}\P_a(x'\mid x)v^*(x').
\end{equation*}


A {\em multiagent MDP} (MMDP) is an extension of MDPs to multiagent settings and can be described as a tuple 
\begin{equation*}
    \M=(N,\X,\set{\A^n,n=1,\ldots,N},\set{\P_{\vec{a}},\vec{a}\in\A},r,\gamma),   
\end{equation*}
where $N$ is the number of agents, $\A^n$ is the individual action space for agent $n$, and $\P_{\vec{a}}$ is the transition probability matrix associated with joint action $\vec{a}$. We write $\A$ to denote the set of all joint actions, corresponding to the Cartesian product of all individual action spaces $\A^n$. We also denote an element of $\A^n$ as $a^n$ and an element of $\A$ as a tuple $\vec{a}=(a^1,\ldots,a^N)$, with $a^n\in\A^n$. We write $\vec{a}^{-n}$ to denote a reduced joint action, i.e., a tuple $\vec{a}^{-n}=(a^1,\ldots,a^{n-1},a^{n+1},\ldots,a^N)$, and thus $\A^{-n}$ is the set of all reduced joint actions. We adopt, for policies, a similar notation. Specifically, we write $\pi^n$ to denote an individual policy for agent $n$, $\vec{\pi}=(\pi^1,\ldots,\pi^N)$ to denote a joint policy, and $\vec{\pi}^{-n}$ to denote a reduced joint policy.

The common goal of the agents in an MMDP is to select a joint policy, $\vec{\pi}^*$, such that $v^{\vec{\pi}^*}(x)\geq v^{\vec{\pi}}(x)$, where, as before, 
\begin{equation}
v^{\vec{\pi}}(x)=\EE[\vec{A}_t\sim\vec{\pi}(X_t)]{\sum_{t=0}^\infty\gamma^tR_t\mid X_0=x}.
\end{equation}
In other words, an MMDP is just an MDP in which the action selection process is distributed across $N$ agents, and can be solved by computing $\vec{\pi}^*$ from $v^*$ as standard MDPs.


Finally, a {\em partially observable MDP}, or POMDP, is an extension of MDPs to partially observable settings. A POMDP can be described as $(\X,\A,\Z,\set{\P_a,a\in\A},\set{\O_a,a\in\A},r,\gamma)$, where $\X$, $\A$, $\set{\P_a,a\in\A}$, $r$, and $\gamma$, are the same as in MDPs, $\Z$ is the {\em observation space}, and $\O_a$ is the observation probability matrix, where
$\O_a(z\mid x)=\PP{Z_{t+1}=z\mid X_{t+1}=x,A_t=a}.$
The {\em belief} at time step $t$ is a distribution $\vec{b}_t$ such that
\begin{equation*}
\begin{split}
b_t(x) \triangleq ~ & \mathbb{P}[X_t=x\mid X_0\sim\vec{b}_0,A_0=a_0, \\ 
& Z_1=z_1,\ldots,Z_t=z_t],
\end{split}
\end{equation*}
where $\vec{b}_0$ is the initial state distribution. Given the action $a_t$ and the observation $z_{t+1}$, we can update the belief $b_t$ to incorporate the new information yielding
\begin{equation}\label{Eq:Belief-update}
\begin{split}
  b_{t+1}(x') 
    & =\bel(b_t,a_t,z_{t+1}) \\
    & \triangleq\frac{1}{\rho_{t+1}}\sum_{x\in\X}b_t(x)\P(x'\mid x,a_t)\O(z_{t+1}\mid x',a_t),
\end{split}
\end{equation}
where $\rho_{t+1}$ is a normalization factor. Every finite POMDP admits an equivalent {\em belief-MDP} with $b_t$ being the state of this new MDP at time step $t$. A policy in a POMDP can thus be seen as mapping $\pi$ from beliefs to distributions over actions, and we define
\begin{equation}
v^\pi(b)\triangleq\EE[A_t\sim\pi(b_t)]{\sum_{t=0}^\infty\gamma^tR_t\mid b_0=b}.
\end{equation}
As in MDPs, the value function associated with an optimal policy is denoted as $v^*$ and can be computed using, for example, point-based approaches \citep{pineau06jair}. From $v^*$, the optimal $Q$-function can now be computed as
\begin{equation*}
\begin{split}
q^*(b,a) & = \sum_{x\in\X}b(x)\Bigg[r(x,a) + \\ 
    & \gamma\sum_{z\in\Z}\sum_{y\in\X}\P(y\mid x,a)\O(z\mid y,a)v^*(\bel(b,a,z))\Bigg],
\end{split}
\end{equation*}
yielding as optimal any policy $\pi^*$ such that $\pi^*(a\mid b)>0$ only if $a\in\argmax_{a\in\A}q^*(b,a)$.


\section{Ad Hoc Teamwork under Partial Observability}


We consider a team of $N$ agents engaged in a cooperative task (henceforth referred as ``target task''), described as an MMDP $m=(N,\X,\set{\A_n},\set{\P_a},r,\gamma)$. One of the agents does not know the task beforehand but must, nevertheless, engage in ad hoc teamwork with the remaining agents to complete the unknown task. We refer to such agent as the ``ad hoc agent'' and denote it as $\alpha$, and refer to the remaining $N-1$ agents collectively as the ``teammates''. Formally, we treat the teammates as a ``meta-agent'' and denote it as $-\alpha$. 

We assume that the teammates all know the target task. The ad hoc agent, however, may not know the target task nor the teammates' policies. Instead, it knows only that the combination between target task and teammate policies is one among $K$, modeled as an MMDP
\begin{equation}
m_k=(2,\X_k,\set{\A^{\alpha},\A_k^{-\alpha}},\set{\P_{k,\vec{a}},\vec{a}\in\A},r_k,\gamma_k).
\end{equation}
Note that we require the action space of the ad hoc agent, $\A^\alpha$, to be the same in all tasks. Other than that, we impose no restrictions on the state space, dynamics, or reward describing these tasks (in particular, they may all be different).

Let $\pi^{-\alpha}_k$ denote a teammates policy for $m_k$, $k=1,\ldots,K$. We have
\begin{equation}\label{Eq:Transitions}
\begin{split}
\P_k(y\mid x,a^\alpha) & \triangleq \mathbb{P}[X_{t+1}=y\mid X_t=x,A^\alpha=a^\alpha, \\
& A^{-\alpha}\sim\pi^{-\alpha}_k(x),M=m_k],
\end{split}
\end{equation}
for $x,y\in\X_k$, where we write $M=m_k$ to denote the fact that the transitions in \eqref{Eq:Transitions} concerns task $k$.

Let us now suppose that, at each moment, the ad hoc agent cannot observe the underlying state of the environment. Instead, at each time step $t$, the agent can only access an observation $Z_t$. We assume that the observations $Z_t$ take values in a (task-independent) set $\Z$ and depend both on the underlying state of the environment and the previous action of the agent (not the teammates). Specifically, for each task $k=1,\ldots,K$, we assume that there is a family of task-dependent observation probability matrices, $\O_{k,a^\alpha},k=1,\ldots,K,a^\alpha\in\A^\alpha$, with
\begin{equation}
\begin{split}
\O_k & (z\mid x,a^\alpha) = \\ 
& \PP{Z_t=z\mid X_t=x,A^{\alpha}_{t-1}=a^\alpha,M=m_k}.
\end{split}
\end{equation}
The elements $[\O_{k,a^\alpha}]_{xz}$ are only defined for $x\in\X_k$. Thus, from the ad hoc agent's perspective, each task $k$ defines a POMDP $\hat{m}_k$ corresponding to the tuple
\begin{equation*}
(\X_k,\A^\alpha,\Z,\set{\P_{k,a^\alpha},a^\alpha\in\A^\alpha},\set{\O_{k,a^\alpha},a^\alpha\in\A^\alpha},r_k,\gamma_k).
\end{equation*}
We denote the solution to $\hat{m}_k$ as $\hat{\pi}_k$.


\subsection{Algorithm}

We adopt a Bayesian framework and treat the target task/teammate model as a random variable, $M$, taking values in the set of possible model descriptions, $\M=\set{m_1,\ldots,m_K}$. For $m_k\in\M$, let $p_0(m_k)$ denote the ad hoc agent's prior over $\M$. Additionally, let $H_t$ denote the random variable corresponding to the history of the agent up to time step $t$, defined as
\begin{equation}
H_t=\set{a^\alpha_0,z_1,a^\alpha_1,z_2,\ldots,a^\alpha_{t-1},z_t}.	
\end{equation}
Then, given a history $h_t$, we define 
\begin{equation}
p_t(m_k)\triangleq\PP{M=m_k\mid H_t=h_t}, \qquad m_k\in\M.
\end{equation}
The distribution $p_t$ corresponds to the agent's belief about the target model at time step $t$. The action for the ad hoc agent at time step $t$ can be computed within our Bayesian setting as
\begin{equation}
\nonumber
\pi_t(a^\alpha\mid h_t)
  =\sum_{k=1}^K\hat{\pi}_k(a^\alpha\mid b_{k,t})p_t(m_k),
\label{Eq:Policy}
\end{equation}
where
\begin{equation}
b_{k,t}(x)\triangleq\PP{X_t=x\mid H_t=h_t,M=m_k},
\end{equation}
for $x\in\X_k$. Upon selecting an action $a^\alpha_t$ and making a new observation $z_{t+1}$, we can update $p_t$ by noting that
\begin{equation*}
\begin{split}
p_{t+1}(m_k) & =\frac{1}{\rho} p_t(m_k) \\ 
& \cdot \PP{A^\alpha_t=a^\alpha_t,Z_{t+1}=z_{t+1}\mid M=m_k,H_t=h_t},
\end{split}
\end{equation*}
where $\rho$ is some normalization constant. Moreover,
\begin{equation*}
\begin{split}
& \PP{A^\alpha_t =a^\alpha_t,Z_{t+1}=z_{t+1}\mid M=m_k,H_t=h_t} \\
&=\sum_{y\in\X_k}\O_k(z_{t+1}\mid y,a^\alpha_t) \pi_t(a^\alpha\mid h_t) \\ 
& \cdot \PP{X_{t+1}=y\mid A^\alpha_t=a^\alpha_t,H_t=h_t,M=m_k}, 
\end{split}
\end{equation*}
where the last equality follows from the fact that the agent's action selection given the history does not depend on the model $M$. Therefore,
\begin{align*}
\lefteqn{\PP{A^\alpha_t=a^\alpha_t,Z_{t+1}=z_{t+1}\mid M=m_k,H_t=h_t}}\\
&=\sum_{x,y\in\X_k}\O_k(z_{t+1}\mid y,a^\alpha_t)
                    \P_k(y\mid x,a^\alpha_t)
                    b_{k,t}(x)\pi_t(a^\alpha\mid h_t).
\end{align*}
Putting everything together, we get
\begin{equation}\label{Eq:Task-update}
\begin{split}
p_{t+1} (m_k) =\frac{1}{\rho}\sum_{x,y\in\X_k} & \O_k(z_{t+1}\mid y,a^\alpha_t) \P_k(y\mid x,a^\alpha_t) \\ 
& \cdot b_{k,t}(x) \pi_t(a^\alpha\mid h_t)p_t(m_k),
\end{split}
\end{equation}
with
\begin{equation*}
\begin{split}
\rho = \sum_{k=1}^K\sum_{x,y\in\X_k} & \O_k(z_{t+1}\mid y,a^\alpha_t) \P_k(y\mid x,a^\alpha_t) \\ 
& \cdot b_{k,t}(x) \pi_t(a^\alpha\mid h_t)p_t(m_k).
\end{split}
\end{equation*}

Note that the update \eqref{Eq:Task-update} requires the ad hoc agent to keep track of the beliefs $b_{k,t}$ for the different POMDPs $\hat{m}_k$. In other words, at each time step $t$, upon executing its individual action $a^\alpha_t$ and observing $z_{t+1}$, the agent updates each belief $b_{k,t}$ using \eqref{Eq:Belief-update}, yielding
\begin{equation}\label{Eq:Belief-update-2}
b_{k,t+1}(y)=\frac{1}{\rho}\sum_{x\in\X_k}b_{t,k}(x)\P_k(y\mid x,a^\alpha_t)\O_k(z_{t+1}\mid y,a^\alpha_t),
\end{equation}
where $\rho$ is the corresponding normalization constant. Since some of the computations in the update \eqref{Eq:Task-update} are common to the update \eqref{Eq:Belief-update-2}, some computational savings can be achieved by caching the intermediate values.

Finally, at each time step $t$, we define the {\em loss} for selecting an action $a^\alpha\in\A^\alpha$ when the target model is $m_k$ to be
\begin{equation}
\ell_t(a^\alpha\mid m_k)=v^{\hat{\pi}_k}(b_{k,t})-q^{\hat{\pi}_k}(b_{k,t},a^\alpha),
\end{equation}
where $\hat{\pi}_k$ is the solution to the POMDP $\hat{m}_k$. 

It is important to note that both $v^{\hat{\pi}_k}(b_{k,t})$ and $q^{\hat{\pi}_k}(b_{k,t},a^\alpha)$ can be computed from $\hat{v}^{\hat{\pi}_k}$ at runtime, while the latter can be computed offline when solving the POMDP $\hat{m}_k$. A bound on $\ell_t(a^\alpha\mid m_k)$ can be found in Section \ref{sec:bound} of the Appendix. Note also that $\ell_t(a^\alpha\mid m_k)\geq 0$ for all $a^\alpha$, and $\ell_t(a^\alpha\mid m_k)=0$ only if $\hat{\pi}_k(a^\alpha\mid b_{k,t})>0$. 

\section{Evaluation}%
\label{Sec:Evaluation}

We setup the ad hoc evaluation method from Stone et al. \citep{stone10aaai}. A single trial therefore consists on running an agent $\alpha_0$ (representing the ad hoc agent) with an agent $\alpha_1$ (representing the teammate) in an environment which is modeled as a POMDP. We then register the total accumulated reward in order to assess the overall team's performance. Since the assumption that there is no available reward signal nor visible teammate's action has never been explored, we compare our approach, ATPO, against five other agents which take on the role of $\alpha_0$:

\begin{itemize}
    
    \item \emph{Value Iteration (non-ad hoc; full observability)}: an agent which is able to perfectly observe the state of the environment and knows the underlying MMDP's optimal policy. Performance reference not as baseline, but as an `oracle' approach for the best theoretical performance in each domain;

    \item \emph{Perseus (non-ad hoc; partial observability)}: an agent, which like ATPO, is only able to observe a partial observation of the environment, but unlike ATPO, knows the task it is performing and the teammate's behavior;

    \item \emph{Random-Picker (ad hoc, partial observability)}: an agent which, like ATPO, also has the library of models and Perseus solutions but selects one randomly in each step (allowing an evaluation of the similarity between possible models;

    \item \emph{BOPA (ad hoc; full observability)}: the current state-of-the-art in ad hoc teamwork which relies on the state of the environment;

    \item \emph{Random (non-ad hoc)}: an agent which selects its action randomly in every step. Reference as the worst possible behavior in each domain;
    
\end{itemize}

We then ask three main research questions:
\begin{enumerate}
    \item Is ATPO effective in assisting the teammates in completing the tasks?
    \item Is ATPO robust in scaling to POMDPs with large state spaces?
    \item Is ATPO robust in scaling to large sets of POMDPs?
\end{enumerate}

\subsection{Domains}

In order to answer all three research questions, we setup a total of 70 POMDPs from 11 different domains. Further details regarding each domain can be found in Appendix~\ref{Sec:Envs}. Table \ref{table:domains} summarizes the characteristics of all POMDPs from each of the eleven domains. Some domains allow for different teammates (varying in POMDP's transition probabilities $\P$, which incorporate the policy of the teammate $\pi_1$), others for different task goals (by varying the reward function $r$), and others for both. One domain (abandoned power plant) even allows for different state spaces between different tasks (varying both $r$ and $\X$).

\subsubsection{Gridworld}

The gridworld domain represents a standard multi-agent navigation problem. In the gridworld domain, $N$ agents must reach $N$ static destination (goal cells). Goal cells are fixed and not encoded in the state, only in the reward. Different possible configurations for the goals thus correspond to the different tasks in $\M$.

\subsubsection{Pursuit}

The pursuit domain used in our work is a modification of the original Pursuit domain where we add partial state observability. In it, $N$ agents ($N=2$ in our experiments) must capture a single moving prey by surrounding it in a coordinated way.

\subsubsection{Abandoned Power Plant}

The abandoned power plant domain models a scenario where two agents---a robot and a human---must cooperate in order to secure an abandoned power plant with six rooms, where three are yet to be explored and two are filled with toxic waste. The ad hoc agent takes in the role of the robot and the teammate the role of the human. In this domain, there are two possible tasks which the human may be performing---charting the entire plant (exploring the three unexplored rooms) and securing the entire plant (by cleaning the two dirty rooms of all available toxic material).
Each task has different modelling requirements, and as such, even though both are modeled as POMDPs, they have different state spaces.

\subsubsection{NTU, ISR, MIT, Pentagon and CIT}

The ntu, isr, mit, pentagon and cit domains \citep{melo2009learning, hu2015learning} all model navigation tasks where two agents---$\alpha_0$ and $\alpha_1$ are spawned in random location of a map and must reach two destinations. Unlike the gridworld domain, which models an open gridworld, the ntu, isr, mit, pentagon and cit domains all model close quarters gridworlds, with walls and collision between the agents. The domains abide by the same base rules yet differ in terms of the map layout (with some being smaller and easier to navigate and others being larger and harder to navigate). They therefore have different state spaces, however, with states modeled the same way.

\subsubsection{Overcooked}

The Overcooked domain \citep{carroll19nips} models a scenario where two agents, a helper and a cook, are required to cooperate in order to cook soups. In this case, tasks represent different teammates which the ad hoc agent must (i) identify; and (ii) adapt to, in order to cook soups. Our ad hoc agent plays the role of the {\em helper}, which has the goal of providing the cook with onions and plates, placing them on a kitchen counter and the teammate plays the role of the {\em cook}, which has the goal of cooking a soup using three onions and serving it in a plate (which is then dispatched through a window).

\begin{table*}[p]
\centering
\caption{Multi-agent domains used for the ad hoc evaluation. $\pi_1$ represents whether the domain allows for varying the teammate's policy, $r$ represents whether the domain allows for varying the target task, $|\X|, |\A|$ and $|\Z|$ represent the total number of states, actions and observations per POMDP from each domain (respectively), $\epsilon$ represents the amount of noise used in computing the transition and observation probabilities for each POMDP, $h$ represents the number of steps in the trajectory horizon during which each POMDP is ran as an environment and, finally, $|\M|$ represents the total number of different setup POMDPs for each domain (which also corresponds to the size of our approach's model library).}
\begin{tabular}{lcccccccc}

    \toprule
    \bf Environment & \bf $\pi_1$ & \bf $r$ & \bf $|\X|$ & \bf $|\A|$ & \bf $|\Z|$ & \bf $\epsilon$ & \bf $h$ & \bf $|\M|$ \\

    \midrule

    gridworld & non-varying & varying & 626 & 5 & 81 & 0.20 & 50 & $2\to32$ \\
    pursuit-task & non-varying & varying & 626 & 5 & 81 & 0.20 & 75 & 4 \\
    pursuit-teammate & varying & non-varying & 626 & 5 & 81 & 0.15 & 85 & 2 \\
    pursuit-both & varying & varying & 626 & 5 & 81 & 0.15 & 85 & 8 \\
    abandoned power plant & non-varying & varying & varying (97, 105) & 6 & 6 & 0.20 & 50 & 2 \\
    ntu & non-varying & varying & 241 & 5 & 81 & 0.20 & 75 & 4 \\
    overcooked & varying & non-varying & 1730 & 4 & 1730 & 0.00 & 50 & 4 \\
    isr & non-varying & varying & 1807 & 5 & 81 & 0.20 & 75 & 3 \\
    mit & non-varying & varying & 2163 & 5 & 81 & 0.20 & 75 & 3 \\
    pentagon & non-varying & varying & 2653 & 5 & 81 & 0.20 & 75 & 3 \\
    cit & non-varying & varying & 4831 & 5 & 81 & 0.10 & 85 & 3 \\
    
    \bottomrule
\end{tabular}
\label{table:domains}
\end{table*}

Given that we are evaluating a total of six agents in the role of $\alpha_0$ and three of them (ATPO, Perseus, and Random-Picker) require the approximate solutions to the POMDPs, we solve all 70 POMDPs from each domain using the Perseus algorithm \citep{spaan2005perseus}. Hyperparameters for the Perseus algorithm can be found in Table~\ref{table:hyperparameters} of the Appendix and the times it takes to both setup and solve each POMDP can be found in Table~\ref{table:times}.

\begin{table*}[p]
    \centering
    \caption{Times to setup and solve POMDPs from each domain. Setup times represent the total time it takes to create the data structures of the POMDP and solve times represent the total time it takes the Perseus algorithm \citep{spaan2005perseus} to obtain a policy.}
    \begin{tabular}{lrlrl}
    
        \toprule
        
        \bf Environment & \bf Avg. Time to Setup & (Std. Dev.) \bf & \bf Avg. Time to Solve & (Std. Dev.)\\
        
        \midrule
        gridworld & 27s 411ms & (±668ms) & 41m 02s 934ms & (±41m 10s 857ms) \\
        pursuit-task & 03m 38s 023ms & (±02s 316ms) & 2h 30m 54s 820ms & (±1h 03m 45s 600ms) \\
        pursuit-teammate & 03m 21s 236ms & (±112ms) & 2h 31m 23s 526ms & (±14m 45s 856ms) \\
        pursuit-both & 03m 31s 369ms & (±03s 085ms) & 3h 03m 02s 597ms & (±1h 16m 14s 350ms) \\
        abandoned power plant & 01s 406ms & (±136ms) & 43s 322ms & (±19s 601ms) \\
        ntu & 14s 004ms & (±502ms) & 10m 05s 246ms & (±01m 24s 147ms) \\
        overcooked & 04m 00s 960ms & (±19s 409ms) & 05m 22s 543ms & (±01m 16s 879ms) \\
        isr & 08m 22s 774ms & (±19s 985ms) & 6h 02m 58s 980ms & (±1h 15m 47s 677ms) \\
        mit & 11m 41s 286ms & (±52s 980ms) & 5h 43m 18s 890ms & (±4h 20m 11s 880ms) \\
        pentagon & 16m 14s 605ms & (±13s 107ms) & 4h 55m 49s 658ms & (±1h 31m 04s 993ms) \\
        cit & 51m 45s 357ms & (±58s 266ms) & 11h 53m 30s 001ms & (±5h 06m 50s 992ms) \\
    
        \bottomrule
        
    \end{tabular}
    \label{table:times}
\end{table*}

\subsection{Results}

We run, for all 11 environments, 32 independent trials for each of the six agents in the role of $\alpha_0$. In each trial, we randomly select one of the possible POMDPs $m \in \M$ and run the interaction for a horizon of $h$ timesteps (the same horizon used for the Perseus algorithm, which can be found in Table~\ref{table:hyperparameters} of the Appendix). We then compare the average accumulated reward over the $h$ timesteps and 32 trials for all agents. Table~\ref{table:results} displays the obtained results.

\begin{table*}[p]
    \centering
    \caption{Average accumulated reward for the six agents in the eleven domains over 32 trials. BOPA requires a common state-space, so it cannot be run in the abandoned power plant scenario.}
    \resizebox{\textwidth}{!}{%
    \begin{tabular}{lrlrlrlrlrlrl}
    
        \toprule
        
        \bf Environment & 
        \multicolumn{2}{c}{\bf Value Iteration} & 
        \multicolumn{2}{c}{\bf Perseus} & 
        \multicolumn{2}{c}{\bf ATPO (ours)} & 
        \multicolumn{2}{c}{\bf Random-Picker} &
        \multicolumn{2}{c}{\bf BOPA} &
        \multicolumn{2}{c}{\bf Random} \\
        \midrule
        
        gridworld & 95.62 & (±1.54) & 93.44 & (±3.13) & 86.53 & (±9.12) & 39.91 & (±61.34) & 93.38 & (±3.14) & 5.88 & (±64.13) \\
        pursuit-both & 90.19 & (±8.32) & 84.44 & (±11.76) & 82.72 & (±12.81) & 35.34 & (±63.18) & 90.06 & (±7.47) & 20.72 & (±69.47) \\
        pursuit-task & 92.72 & (±4.87) & 88.03 & (±7.51) & 82.59 & (±11.18) & 47.59 & (±55.46) & 92.81 & (±3.48) & 14.59 & (±71.34) \\
        pursuit-teammate & 89.88 & (±8.38) & 86.97 & (±12.04) & 86.41 & (±14.52) & 81.34 & (±31.56) & 91.38 & (±6.13) & -7.66 & (±78.7) \\
        abandoned power plant & 94.91 & (±1.99) & 95.12 & (±2.1) & 94.75 & (±2.35) & 93.81 & (±4.97) & N.A. & N.A. & 47.44 & (±62.31) \\
        ntu & 98.00 & (±0.0) & 97.25 & (±0.43) & 97.19 & (±0.53) & 97.28 & (±0.51) & 98.0 & (±0.0) & 88.00 & (±7.42) \\
        overcooked & 39.12 & (±60.73) & 39.00 & (±60.63) & 25.31 & (±66.43) & 27.94 & (±64.58) & 24.81 & (±66.19) & -46.75 & (±18.1) \\
        isr & 92.41 & (±2.68) & 91.16 & (±4.68) & 82.62 & (±12.19) & 60.59 & (±59.15) & 91.66 & (±4.22) & -23.31 & (±68.05) \\
        mit & 84.50 & (±0.87) & 84.09 & (±0.58) & 83.72 & (±0.57) & 83.75 & (±0.75) & 84.31 & (±0.88) & -71.41 & (±20.01) \\
        pentagon & 91.00 & (±0.0) & 90.62 & (±0.6) & 81.19 & (±11.04) & 53.28 & (±56.53) & 85.5 & (±6.71) & -45.94 & (±55.14) \\
        cit & 86.22 & (±2.27) & 85.09 & (±2.53) & 75.66 & (±9.52) & 69.91 & (±14.46) & 84.88 & (±2.94) & -81.25 & (±20.88) \\

        \bottomrule
        
    \end{tabular}}
    \label{table:results}
\end{table*}

\begin{figure*}[!tb]
    \centering
    \includegraphics[width=.75\textwidth]{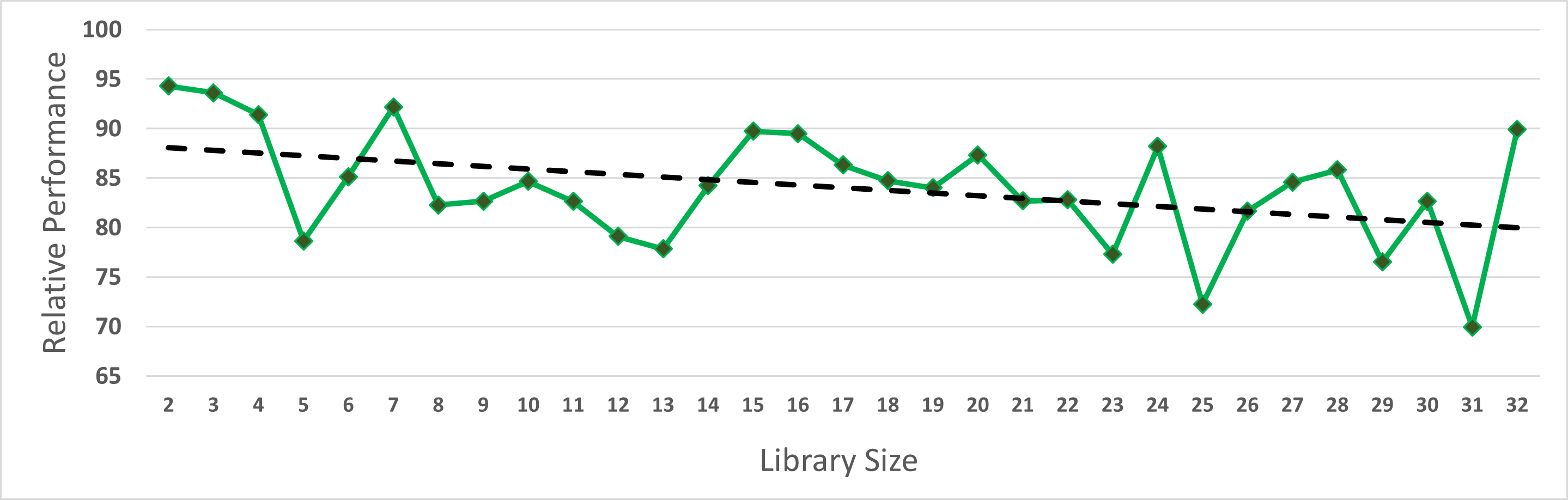}
    \caption{Varying the total number of possible POMDPs from the gridworld domain in ATPO's library. Results correspond to the relative performance, normalized between the best possible performance (100\%, obtained by the Value Iteration agent) and worst possible performance (0\%, obtained by the Random Agent).}
    \label{fig:scalability-library}
\end{figure*}

At first glance, we may from these results conclude which domains are easier and which domains are harder. This can be seen by looking at the performance obtained by the Random Agent. Domains such as ntu and abandoned power plant, in which the Random Agent was able to achieve its highest possible results, are also the domains with the smallest state space. On the other hand, domains like cit and mit, where the Random Agent obtained its worst results, are also the domains with the highest number of possible states, showcasing that the problem difficulty is associated with the number of states in the POMDP.

From these results we are able to observe that all domains were effectively solved by the Perseus algorithm, providing a solid foundation for ATPO's library. In one domain (abandoned power plant, one of the easiest), the agent which knows the task and teammate behavior, but is unable to observe the state of the environment and therefore resorts the Perseus policy (Perseus agent), was even able to, on average, slightly outperform the Value Iteration agent (although an insignificant difference).

Given that ATPO is required to identify, within its library $\M$, the POMDP which best models the environment (unlike the Perseus agent which already knows which one it is), we expected it to underperform the Perseus agent. In all domains this phenomenon was observed, further showcasing that our testbed is, in fact, a fair testbed for ATPO's evaluation.

We can also compare ATPO against the Random-Picker agent in order to identify which environments contain the most similar tasks and teammates. Given that the Random-Picker can be seen as an instance of ATPO with a constant uniform belief over the possible models, it is expected in most cases to underperform ATPO, since Random-Picker will most likely use the wrong policy. In domains where this doesn't happen, and Random-Picker is able to obtain similar results to ATPO, it means the tasks and teammates are very similar. As expected, ATPO outperforms Random-Picker in the different environments, showcasing the impact of the task identification ability of ATPO. In some scenarios, the difference is not very large (e.g., MIT, NTU), indicating that the task identification component is not critical to attain a good performance. However, in several other tasks, the difference is quite significant (e.g., Gridworld, Pentagon, Pursuit), and clearly shows the advantages of ATPO. As for BOPA, since the agent has full state observability, it is unsurprising that it attains better performance than ATPO - the difference attesting, to some extent, the impact that partial state observability can have in the ability of the agent to act.

Regarding the impact of the ATPO's Bayesian inference, our experiments already hint to the impact of model identification in the performance of ATPO. However, we should note that during our experiments we also keep track of the beliefs over tasks, as ATPO interacts with the environment. By analyzing how the beliefs evolve over time, we can highlight a couple of more salient aspects. Specifically, if we look at the average probability associated to the correct model against that associated with the other tasks, we note that i) the correct task is identified, in average, after $5$ timesteps, ii) in average, after $10$ timesteps, the correct task has an associated probability of around $70$\% and iii) in average, after $20$ timesteps, the correct task has an associated probability of around $90$\%. These values are not uniform across domains, but roughly indicate that the proposed approach is indeed able to effectively “zero-in” on the correct task. Analyzing this same results in more detail, now considering how the beliefs evolve for each scenario individually, we can report that, in the most difficult domains, it takes around $10$-$15$ timesteps for the correct task to reach the highest probability (becoming the most likely one), and by the end of the interaction, ATPO had near $100$\% certainty it was in fact the correct one. In the smaller and easier domains, it took only around $3$-$4$ timesteps for the correct task to be the most likely one and, by the end of the interaction, ATPO also reached around $100$\% certainty that it was the correct one. A visualization of the average entropy of the beliefs for ATPO and BOPA can be found in Figure~\ref{fig:entropy} in the Appendix.

Moving on, we can treat the results obtained by the Random Agent as a lower bound to an agent's performance and the performance of the Value Iteration agent (which knows the task being performed, the behavior of the teammate, and the current state of the environment) as an upper bound. If we normalize our results into a relative scale (where $100\%$ corresponds to the performance of the Value Iteration agent and $0\%$ corresponds to the performance of the Random agent) we are now able to more effectively evaluate our approach. Table~\ref{table:results-relative} displays the same results from Table~\ref{table:results}, but normalized.

\begin{table}
    \centering
    \caption{ATPO's results from Table~\ref{table:results} normalized between the performance obtained by the Value Iteration agent (100\%) and performance obtained by the Random Agent (0\%).}

    \resizebox{0.5\linewidth}{!}{%
    \begin{tabular}{lr}
    
        \toprule
        
        \bf Environment & \bf ATPO \\
        \midrule
        
        gridworld             & 89.87\% \\
        pursuit-both          & 89.25\% \\
        pursuit-task          & 87.03\% \\
        pursuit-teammate      & 96.44\% \\
        abandoned power plant & 99.66\% \\
        ntu                   & 91.90\% \\
        overcooked            & 83.92\% \\
        isr                   & 91.54\% \\
        mit                   & 99.50\% \\
        pentagon              & 92.84\% \\
        cit                   & 93.69\% \\
        \bottomrule
    
    \end{tabular}}
    \label{table:results-relative}
\end{table}

Having now established a solid evaluation testbed, we can start looking at our approach's results (ATPO). Overall, ATPO was able to obtain above $83.92\%$ performance on all domains (showcasing its effectiveness). Its worst results were on the overcooked domain (with $83.92\%$ relative performance) and its best results were on the abandoned power plant and mit domains (with $99.66\%$ and $99.50\%$, respectively). These results therefore showcase that our approach is effective in assisting unknown teammates in solving unknown tasks resorting only to a partial observation of the environment, and robust in adapting to domains with large state spaces.

\subsection{Increasing ATPO's Library}

We now evaluate ATPO's robustness in scaling to larger libraries of possible POMDPs (our third and final research question). We resort to a single domain---the gridworld domain---and conduct an experiment where we increase the number of POMDPs in ATPO's library $|\M|$ from 2 to 32. We then run, per each of the 31 library sizes, 32 independent trials on the environment (adding up to a total of 992 additional trials). Figure~\ref{fig:scalability-library} reports our obtained relative results (normalized between the best possible performance and worst possible performance, taken from the results obtained in Table~\ref{table:results} by the Value Iteration and Random Agent, respectively).

As we can see from these results, as we increase the number of possible POMDPs in ATPO's library, its performance follows a small downward trend. This is expected, given that the more possible POMDPs there might be to explain the environment, the harder it will be to identify the correct one. Nevertheless, ATPO's performance with up to 32 possible models was always above 70\%, strongly showcasing our approach's robustness in scaling to a higher number of possible models.


\section{Conclusion}

This paper makes three main contributions to the ad hoc teamwork literature: (i) formalizes the setting of ad hoc teamwork under partial observability using a Markovian framework, a setting where an ad hoc agent is unable observe the full state of the environment and the actions of its teammates, (ii) proposes a first-principled, novel model-based approach for this setting and (iii) showcases its effectiveness by conducting an empirical evaluation in a total of 70 POMDPs from 11 different domains. Our results show that our approach (ATPO), relying only on its model library of possible POMDPs describing the multi-agent systems and their respective solutions and partial observations, was able to i) effectively assist unknown teammates in performing unknown tasks, ii) robustly scale to environments with large state spaces (up to 4831 states), iii) robustly scale to a high number of possible models (up to 32 possible POMDPs) and iv) identify which model better describes the environment it is interacting with alongside its respective teammates.

A logical future line of work will be to expand our approach to domains with continuous or arbitrarily large state spaces, requiring ATPO to be adapted using function approximation solutions such as feedforward recurrent neural networks, combined with model-based planning approaches such as MuZero \citep{schrittwieser2020mastering}.

\section*{Acknowledgments}

This work was partially supported by national funds through FCT, Funda\c{c}\~{a}o para a Ci\^{e}ncia e a Tecnologia, under project UIDB/50021/2020 (INESC-ID multi-annual funding), the HOTSPOT project, with reference PTDC/CCI-COM/7203/2020 and the RELEVANT project, with reference PTDC/CCI-COM/5060/2021. In addition, this material is based upon work supported by the Air Force Office of Scientific Research under award number FA9550-22-1-0475, and by TAILOR, a project funded by EU Horizon 2020 research and innovation programme under GA No 952215. The first author acknowledges the PhD grant 2020.05151.BD from FCT. The authors thankfully acknowledge the comments from the anonymous reviewers.

\bibliography{ecai.bib}

\clearpage

\appendix

\section{Loss bounds for ATPO}
\label{sec:bound}

This section provides a bound on the loss incurred by our approach. The bound can be derived from a standard compression lemma from Banerjee \citep{banerjee06icml}, by comparing the performance of an agent using an arbitrary constant distribution over tasks. Let $p$ denote a distribution over $\M$, and define
\begin{align*}
L_t(p)
  &=\sum_{k=1}^Kp(m_k)\ell_t(\hat{\pi}_k\mid m^*),
\end{align*}
where $m^*$ is the (unknown) target task, and
\begin{equation*}
\ell_t(\hat{\pi}_k\mid m^*)=\sum_{a^\alpha\in\A^\alpha}\hat{\pi}_k(a^\alpha\mid b_{k,t})\ell_t(a^\alpha\mid m^*).
\end{equation*}
We have the following result, the proof of which is provided in the supplementary material.

\begin{proposition}\label{Prop:Bound}
Let $q$ denote an arbitrary stationary distribution over the tasks in $\M$, and $\set{p_t}$ the sequence of beliefs over tasks generated by ATPO. Then,
\begin{equation}
\label{Eq:Bound}
\begin{split}
\sum_{t=0}^{T-1}L_t(p_t) & \leq \sum_{t=0}^{T-1}L_t(q)+\sqrt{\frac{2}{T}}\sum_{t=0}^{T-1}\KL(q\parallel p_t) \\ 
    &+\sqrt{\frac{T}{2}}\cdot \frac{R_{\max}^2}{(1-\gamma)^2}.
\end{split}
\end{equation}
\end{proposition}

Unsurprisingly---and aside from the term $\sqrt{\frac{T}{2}}\cdot \frac{R_{\max}^2}{(1-\gamma)^2}$, which is independent of $q$ and grows sublinearly with $T$---the bound in \eqref{Eq:Bound} states that the difference between the performance of ATPO and that obtained using a constant distribution (for example, the distribution concentrated on $m^*$) is similar to those reported by Banerjee \citep{banerjee06icml} for Bayesian online prediction with bounded loss, noting that 
\begin{equation}
\sum_{t=0}^{T-1}\KL(q\parallel p_t)=\KL(\vec{q}\parallel\vec{p}_{0:T-1}),
\end{equation}
where we write $\vec{q}$ and $\vec{p}_{0:T-1}$ refer to distributions over sequences in $\M^T$.

\subsection{Proof of Proposition~\ref{Prop:Bound}}

We use the following compression lemma from Banerjee \citep{banerjee06icml}.

\begin{lemma}\label{Lemma:Compression}
Given a set of hypothesis $\H=\set{1,\ldots,H}$, for any measurable function $\phi:\H\to\IR$ and any distributions $p$ and $q$ on $\H$,
\begin{equation}
\EE[h\sim q]{\phi(h)}-\log\EE[h\sim p]{\exp(\phi(h))}\leq\KL(q\parallel p).
\end{equation}
\end{lemma}

We want to bound the loss incurred by our agent after $T$ time steps. Before introducing our result, we require some auxiliary notation. Let $m^*$ denote the (unknown) target task at time step $t$. The expected loss of our agent at time step $t$ is given by
\begin{equation*}
\begin{split}
L_t(\pi_t)
  &=\EE{\ell_t(A^\alpha\mid m^*)} \\
  &=\sum_{a^\alpha\in\A^\alpha}\pi_t(a^\alpha)\ell_t(a^\alpha\mid m^*) \\
  &=\sum_{k=1}^Kp_t(m_k)\sum_{a^\alpha\in\A^\alpha}\hat{\pi}_k(a^\alpha\mid b_{k,t})\ell_t(a^\alpha\mid m^*)\\
  &=\sum_{k=1}^Kp_t(m_k)\ell_t(\hat{\pi}_k\mid m^*),
\end{split}
\end{equation*}
where, for compactness, we wrote
\begin{equation}
\ell_t(\hat{\pi}_k\mid m^*)=\sum_{a^\alpha\in\A^\alpha}\hat{\pi}_k(a^\alpha\mid b_{k,t})\ell_t(a^\alpha\mid m^*).
\end{equation}

Let $q$ denote an arbitrary distribution over $\M$, and define
\begin{equation}
L_t(q)=\sum_{k=1}^Kq(m_k)\ell_t(\hat{\pi}_k\mid m^*).
\end{equation}
Then, setting $\phi(m_k)=-\eta\ell_t(\hat{\pi}_{k_t}\mid m^*)$, for some $\eta>0$, and using Lemma~\ref{Lemma:Compression}, we have that
\begin{equation}
\EE[m\sim q]{\phi(m)}-\log\EE[m\sim p_t]{\exp(\phi(m))}\leq\KL(q\parallel p_t)
\end{equation}
which is equivalent to
\begin{equation}\label{Eq:Ineq1}
-\log\EE[m\sim p_t]{\exp(\phi(m))}\leq \eta L_t(q)+\KL(q\parallel p_t).
\end{equation}
Noting that $-2\eta\frac{R_{\max}}{1-\gamma}\leq\phi(m)\leq 0$ and using Hoeffding's Lemma,%
\footnote{Hoeffding's lemma states that, given a real-valued random variable $X$, where $a\leq X\leq b$ almost surely, 
\begin{equation}
\EE{e^{\lambda X}}\leq\exp\left(\lambda\EE{X}+\frac{\lambda^2(b-a)^2}{8}\right),
\end{equation}
for any $\lambda\in\IR$. 
}
we have that
\begin{equation}\label{Eq:Ineq2}
-\log\EE[m\sim p_t]{\exp(\phi(m))}\geq\eta L_t(p_t)-\frac{\eta^2R_{\max}^2}{2(1-\gamma)^2}.
\end{equation}
Combining \eqref{Eq:Ineq1} and \eqref{Eq:Ineq2}, yields
\begin{equation}
L_t(p_t)\leq L_t(q)+\frac{1}{\eta}\KL(q\parallel p_t)+\frac{\eta R_{\max}^2}{2(1-\gamma)^2}
\end{equation}
which, summing for all $t$, yields
\begin{equation}
\sum_{t=0}^{T-1}L_t(p_t)\leq \sum_{t=0}^{T-1}L_t(q)+\frac{1}{\eta}\sum_{t=0}^{T-1}\KL(q\parallel p_t)+\frac{T\eta R_{\max}^2}{2(1-\gamma)^2}.
\end{equation}
Since $\eta$ can be selected arbitrarily, setting $\eta=\sqrt{\frac{T}{2}}$ we finally get
\begin{equation}
\begin{split}
\sum_{t=0}^{T-1}L_t(p_t) & \leq \sum_{t=0}^{T-1}L_t(q)+\sqrt{\frac{2}{T}}\sum_{t=0}^{T-1}\KL(q\parallel p_t) \\ 
    & +\sqrt{\frac{T}{2}}\cdot \frac{R_{\max}^2}{(1-\gamma)^2}.
\end{split}
\end{equation}

\section{Domain Descriptions}%
\label{Sec:Envs}

We now provide detailed descriptions of our three test scenarios.

\subsection{Gridworld}

\begin{figure}[!tb]
    \centering
    \includegraphics[width=.33\linewidth]{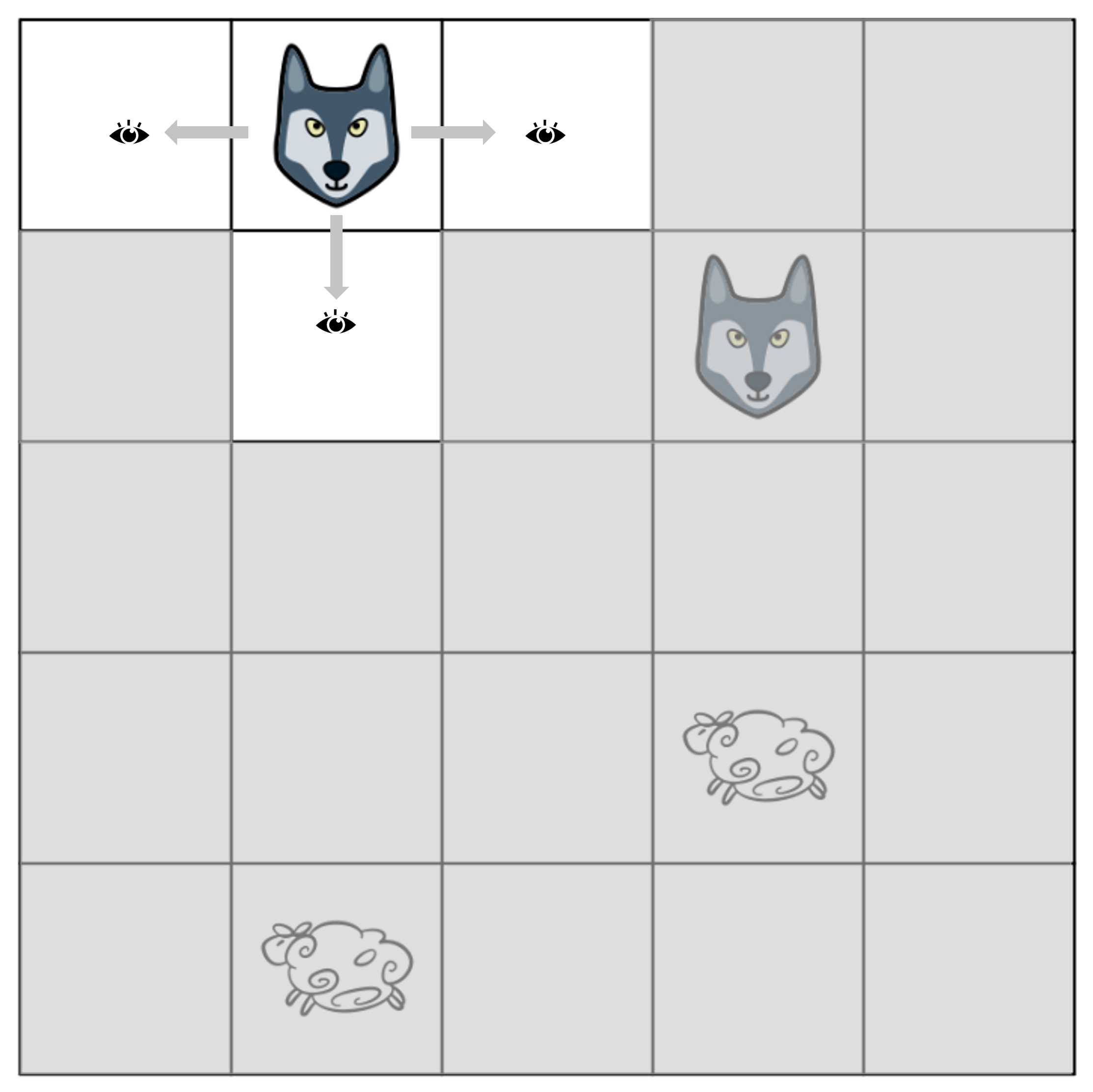}
    \caption{The gridworld domain. Two agents must each navigate to two goal cells.}
    \label{fig:gridworld}
\end{figure}

In the gridworld domain (Fig. \ref{fig:gridworld}), the ad hoc agent can move up, down, left and right, or stay in its current cell. The teammate follows the shortest path to its closest goal. Each moving action succeeds with probability $1-\eps$, for some $\eps\geq 0$, except if there is a wall in the corresponding direction, in which case the position of the agent remains unchanged. When an action fails, the position of the agent remains unchanged.

The ad hoc agent can only observe the neighboring cells. {\em It cannot observe its current position}. Observations are also not flawless: whenever an element (teammate, wall) is in a neighboring cell, there is a probability $\eps$ that the agent will fail to observe it.

We model each possible model $m_k\in\M$ as a POMDP 
\begin{equation*}
(\X,\A^\alpha,\Z,\set{\P_{a^\alpha},a^\alpha\in\A^\alpha},\set{\O_{a^\alpha},a^\alpha\in\A^\alpha},r_k,\gamma).	
\end{equation*}

A state $x\in\X$ contains information regarding the positions of both agents, and is therefore represented as a tuple $x=(c_1, r_1, c_2, r_2)$, where $(c_n, r_n)$ represents the cell (column and row) where agent $n$ is located. 

Each observation $z\in\Z$ are also represented by a tuple $z=(\hat{u}, \hat{d}, \hat{l}, \hat{r})$, where each entry represents what is observed, respectively, above, below, to the left, and to the right of the agent. For each entry there are three possible values: \textit{Nothing}, \textit{Teammate}, \textit{Wall}. The action space $\A^\alpha$ contains five possible actions, \textit{Up, Down, Left, Right} and \textit{Stay}. The transition probabilities $\set{\P_{a^\alpha},a^\alpha\in\A^\alpha}$ map a state $x$ and action $a^\alpha$ to every possible next state $x'$, taking into account the probability $\epsilon$ of the agent failing to move (considering that the teammate actions always succeed and move the teammate towards its closest goal cell). Similarly, the observation probabilities $\set{\O_a^\alpha,a^\alpha\in\A^\alpha}$ map a state $x$ and previous action $a^\alpha$ to every an observation $z$, taking into account the probability $\epsilon$ of the agent failing to observe any given element. The reward function $r_k$ assigns the reward of $-1$ for all time steps except those where both destination have been reached, in which case it assigns a reward of $100$ (and an absorbent state assigned a reward of $0$ afterwards). In other words, since where dealing with an horizon of 50 steps, we do not reset the environment whenever the goals have been reached. Finally, we consider a discount factor $\gamma=0.95$.

\subsection{Pursuit}

\begin{figure}[!tb]
    \centering
    \includegraphics[width=.33\linewidth]{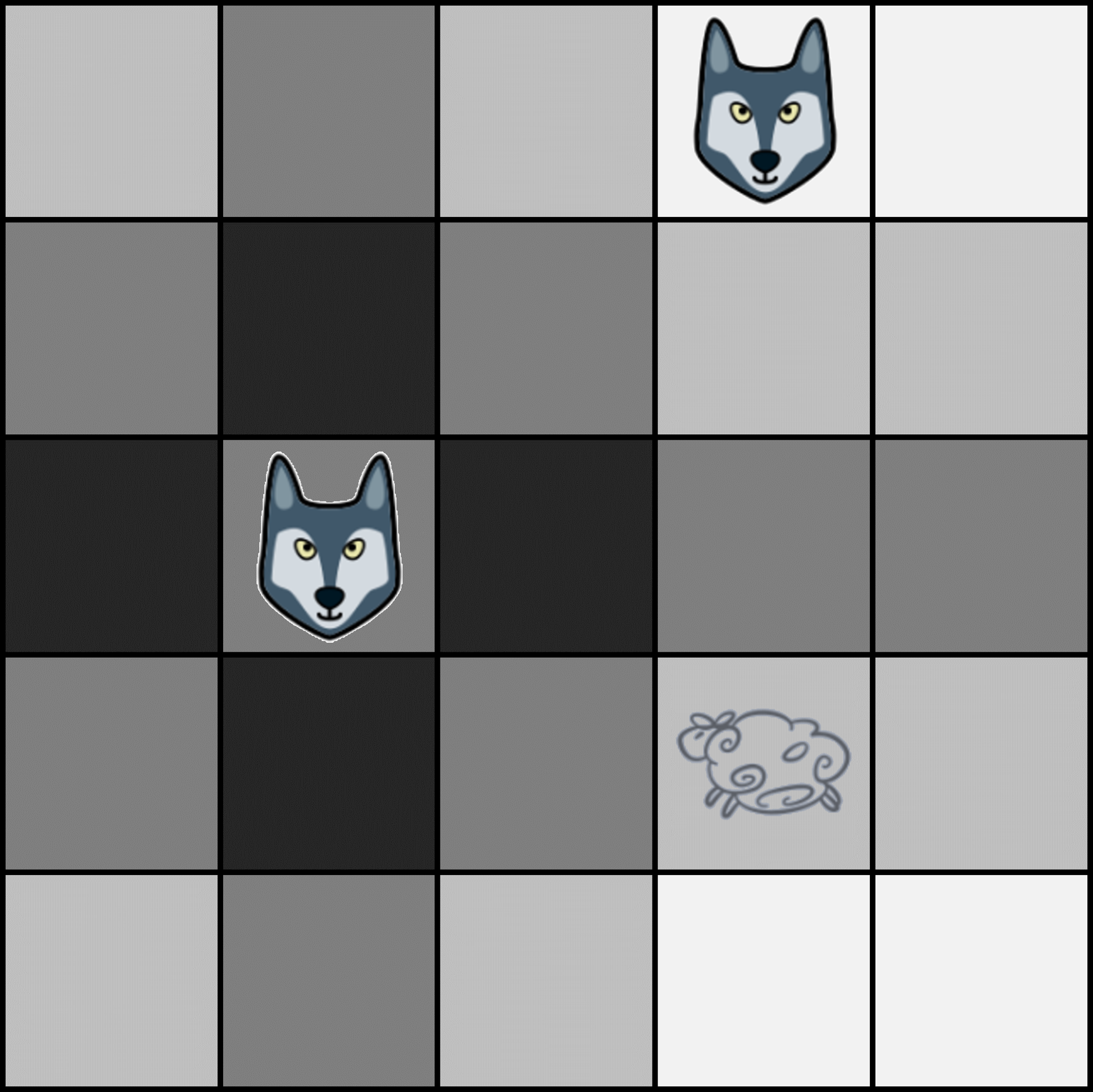}
    \caption{The pursuit domain. Two agents must capture a moving prey.}
    \label{fig:pursuit}
\end{figure}

In the pursuit domain (Fig. \ref{fig:pursuit}), the ad hoc agent is able to observe the elements located in its neighborhood. Specifically, if either the teammate or the prey are located in one of the neighbouring nine cells surrounding the agent, it may be able to observe them with probability $1-\epsilon$, where $\epsilon$ represents the probability of failing the observation.

We model each possible model $m_k\in\M$ as a POMDP 
\begin{equation*}
(\X,\A^\alpha,\Z,\set{\P_{k,a^\alpha},a^\alpha\in\A^\alpha},\set{\O_{a^\alpha},a^\alpha\in\A^\alpha},r_k,\gamma).	
\end{equation*}
Different prey capture configurations vary the reward function $r$ and different teammate policies vary the transition probabilities $\P$. We consider four different capture combinations: (1) \textit{north of prey + south of prey}; (2) \textit{west of prey + east of prey}; (3) \textit{southwest of prey +  northeast of prey}; and (4) \textit{northwest of prey + southeast of prey}. The capture positions may be used interchangeably between the agents (i.e., either the ad hoc agent or the teammate may occupy each of the two required positions). We also consider two possible teammate policies: (1) greedy policy; and (2) teammate aware policy. The greedy policy follows the shortest path to its closest capture position, not taking into account the position of the ad hoc agent and therefore not acting as efficiently as possible, while the teammate aware policy runs an A\* search taking into account the position of the teammate. Each state $x\in\X$ contain information regarding the relative distances to the teammate and prey and is therefore represented by a tuple $x=(d{^{a_1}}_x, d{^{a_1}}_y, d{^p}_x, d{^p}_y)$, where $d{^{a_1}}_x, d{^{a_1}}_y$ represents the relative distance (in units) to the teammate and $d{^p}_x, d{^p}_y$ represents the relative distance (in units) to the prey. Each observation $z\in\Z$ is also represented as a tuple $z=(\hat{a_1}, \hat{p})$, where $\hat{a_1}$ represents an observation cell identifier of the teammate and $\hat{p}$ represents an observation cell identifier of the prey. A cell identifier is a value between $0 and 8$ which represents the nine surrounding cells of the agent. When an observation is failed or the teammate/prey is not in the nine surrounding cells, the ad hoc agent sees the other agent as it were standing in its own position. 
The action space $\A^\alpha$ contains five possible actions, \textit{Up, Down, Left, Right, Stay}. For each teammate type $k$, the transition probabilities $\set{\P_{k,a^\alpha},a^\alpha\in\A^\alpha}$ map a state $x$ and action $a$ to every possible next state $x'$, taking into account the probability of the teammate executing each possible action on $x$ given their policy for task $k$. Similarly, the observation probabilities $\set{\O_a^\alpha,a^\alpha\in\A}$ map a state $x$ and previous action $a^\alpha$ to every possible observation $z$, taking into account the probability $\epsilon(d)$ of the agent failing to observe the position of the other agents. The reward function $r_k$ assigns the reward of $-1$ for all time steps except those where the prey has been cornered, in which case it assigns a reward of $100$ for the one the prey was cornered in and $0$ afterwards (in other words, since where dealing with a finite horizon, we do not reset the environment whenever the prey have been cornered). Finally, we consider a discount factor $\gamma=0.95$.

\subsection{Abandoned Power Plant}

We model each possible model $m_k\in\M$ of the abandoned power plant domain as a POMDP 
\begin{equation*}
(\X_k,\A^\alpha,\Z,\set{\P_{a^\alpha},a^\alpha\in\A^\alpha},\set{\O_{a^\alpha},a^\alpha\in\A^\alpha},r_k,\gamma).	
\end{equation*}

POMDPs from both tasks vary in both state space $|\X|$ and reward function $r$. In POMDPs from the first task---exploration---each state $x\in\X$ is represented as a tuple $x=(\text{room}_{\text{robot}}, \text{room}_{\text{human}}, e_1, e_2, e_3)$ where $\text{room}_{\text{robot}}$ and $\text{room}_{\text{human}}$ represent the identifier of the room in which the robot and human are located (respectively) and $e_i$ represents the status of room $i$ (explored, unexplored). In POMDPs from the second task---cleanup---each state $x\in\X$ is represented as a tuple $x=(\text{room}_{\text{robot}}, \text{room}_{\text{human}}, d_1, d_2)$, where $\text{room}_{\text{robot}}$ and $\text{room}_{\text{human}}$ represent the identifier of the room in which the robot and human are located (respectively) and $d_i$ represent the status of the room $i$ (dirty, clean). The power plant is modeled as a topological graph with six nodes representing the rooms. The edges between nodes represent the connections between the rooms, and therefore, the action space for both the robot and the human contain four base actions---\textit{Move to lowest-index node, Move to second lowest-index node, Move to third lowest-index node and Stay}. The robot has to itself available two additional actions---\textit{Query human for human location and Query human for own location}. The observation $z\in\Z$ is represented by a a value. Whenever the robot moves (instead of querying the human), it is able to observe, with a probability of failure $\epsilon$, its own location, with $z$ representing the id of the room it is in (and $-1$ when failing). Whenever the robot queries the human for the human's or its own location, the human may or may not reply, also with probability $\epsilon$. If the query is successful, $z$ contains the room of the human or the room of the robot (respectively according to the query type). If the query fails, $z=-1$. Finally, we consider a discount factor $\gamma$ of $0.95$ and the reward function $r$ assigns a value of $-1$ the one where, in the case of exploration tasks, all unexplored rooms have been explored and, in the case of the cleanup tasks, all dirty rooms have been cleaned. Afterwards, an absorbent state is assigned a reward of $0$.

\subsection{NTU, ISR, MIT, PENTAGON and CIT}

\begin{figure}[!tb]
    
    \centering
    \includegraphics[width=.2\textwidth]{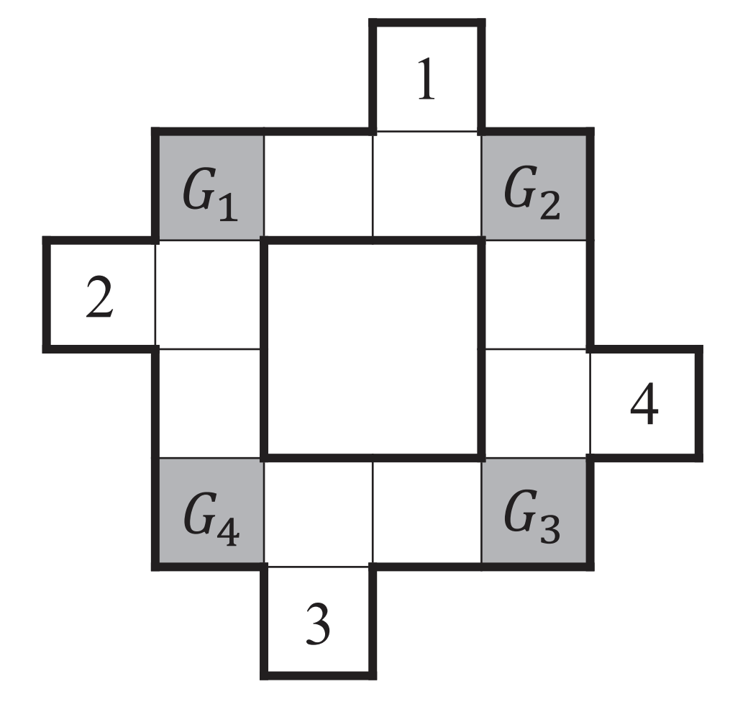}\quad
    \includegraphics[width=.2\textwidth]{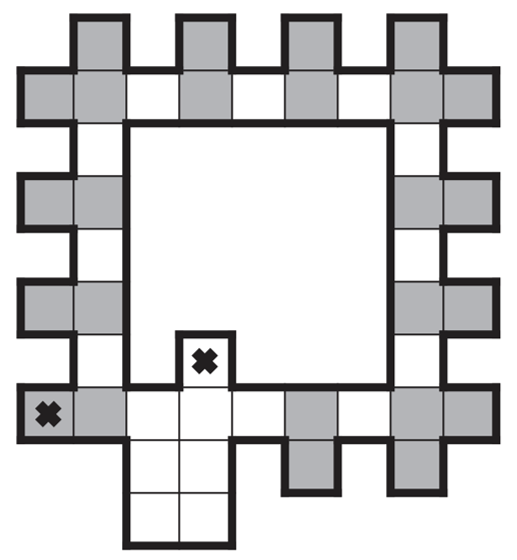}\quad
    \includegraphics[width=.2\textwidth]{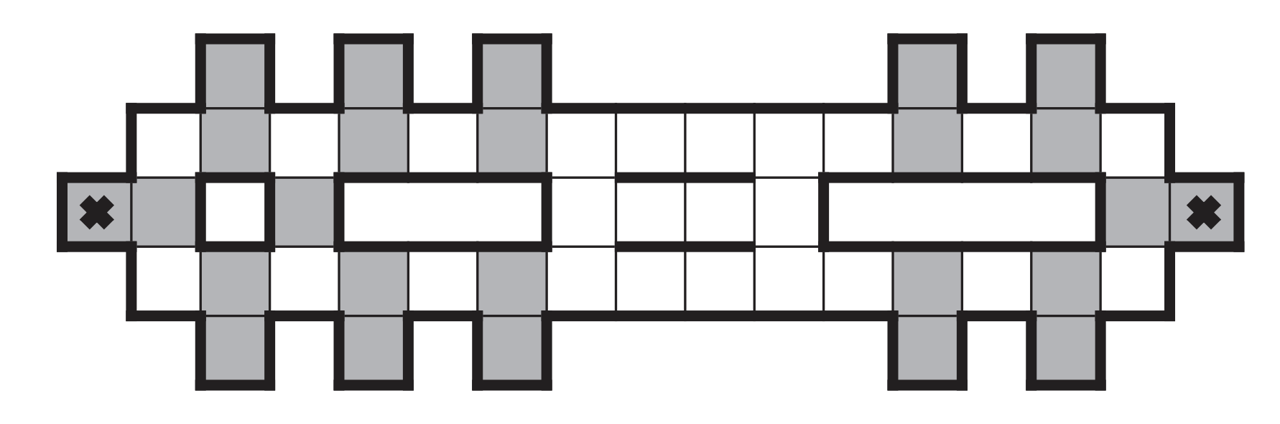}
    
    \medskip
    
    \includegraphics[width=.2\textwidth]{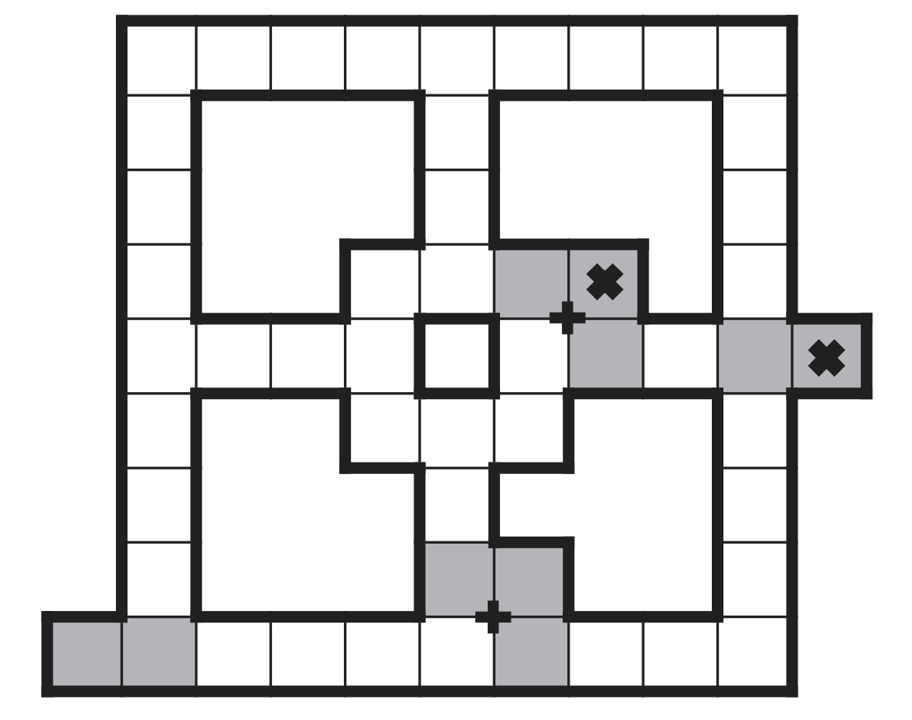}\quad
    \includegraphics[width=.2\textwidth]{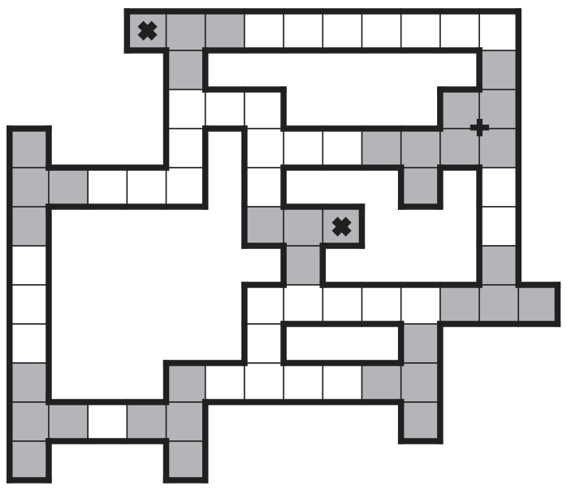}
    
    \caption{The ntu, isr, mit, pentagon and cit domains. Two agents must navigate in close quarters in order to each reach an exit (while taking into account collisions).}
    
    \label{fig:cmu}
\end{figure}

We model each possible model $m_k\in\M$ of the ntu, isr, mit, pentagon and cit domains (Fig. \ref{fig:cmu}) as a POMDP 
\begin{equation*}
(\X,\A^\alpha,\Z,\set{\P_{a^\alpha},a^\alpha\in\A^\alpha},\set{\O_{a^\alpha},a^\alpha\in\A^\alpha},r_k,\gamma).	
\end{equation*}

Different tasks have different goal locations, which translate to different reward functions $r$. A state $x\in\X$ contains information regarding the positions of both agents, and is therefore represented as a tuple $x=(c_1, r_1, c_2, r_2)$, where $(c_n, r_n)$ represents the cell (column and row) where agent $n$ is located. 

Each observation $z\in\Z$ are also represented by a tuple $z=(\hat{u}, \hat{d}, \hat{l}, \hat{r})$, where each entry represents what is observed, respectively, above, below, to the left, and to the right of the agent. For each entry there are three possible values: \textit{Nothing}, \textit{Teammate}, \textit{Wall}. The action space $\A^\alpha$ contains five possible actions, \textit{Up, Down, Left, Right} and \textit{Stay}. The transition probabilities $\set{\P_{a^\alpha},a^\alpha\in\A^\alpha}$ map a state $x$ and action $a^\alpha$ to every possible next state $x'$, taking into account the probability $\epsilon$ of the agent failing to move (considering that the teammate actions always succeed and move the teammate towards its closest goal cell). Similarly, the observation probabilities $\set{\O_a^\alpha,a^\alpha\in\A^\alpha}$ map a state $x$ and previous action $a^\alpha$ to every an observation $z$, taking into account the probability $\epsilon$ of the agent failing to observe any given element. The reward function $r_k$ assigns the reward of $-1$ for all time steps except those where both destination have been reached, in which case it assigns a reward of $100$ (and an absorbent state assigned a reward of $0$ afterwards). In other words, since where dealing with an horizon of 50 steps, we do not reset the environment whenever the goals have been reached. Finally, we consider a discount factor $\gamma=0.95$.

\subsection{Overcooked}

\begin{figure}[!tb]
    \centering
    \includegraphics[width=.33\linewidth]{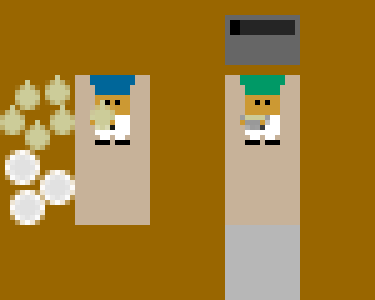}
    \caption{The overcooked domain. Two agents, a cook and an assistant, are required to deliver soups as fast as possible.}
    \label{fig:overcooked}
\end{figure}

We model each possible model $m_k\in\M$ of the Overcooked domain (Fig. \ref{fig:overcooked}) as a POMDP 
\begin{equation*}
(\X,\A^\alpha,\Z,\set{\P_{k,a^\alpha},a^\alpha\in\A^\alpha},\set{\O_{a^\alpha},a^\alpha\in\A^\alpha},r,\gamma).	
\end{equation*}

Varying the teammates varies the transition probabilities. Since the state is fully observable in this domain, observations correspond to the states themselves. Since the policy of the teammate may vary, each different model has distinct transition probabilities. We consider four different cook teammate policies: (1) a teammate which has the optimal policy; (2) a teammate which acts randomly; (3) a teammate which has the optimal policy but always stands near the bottom balcony when receiving ingredients and plates; and (4) a teammate which has the optimal policy but always stands near the upper balcony when receiving ingredients and plates. Each state $x\in\X$ is represented as a tuple $x=(p_a, p_c, h_a, h_c, tb, bb, s)$, where, respectively, $(p_a, p_c)$ represents the cell (top or bottom) and $(h_a, h_c)$ represents the objects (nothing, onion, plate or soup) in hand of the helper and the cook. $(tb, bb)$ represents the objects on the top kitchen counter balcony and bottom kitchen counter balcony (respectively). Finally, $s$ represents the contents of the soup pan (empty, one onion, two onions, cooked soup). The action space $\A$ contains four possible actions, \textit{Up, Down, Noop} and \textit{Act}. The transition probabilities $\set{\P_{k,a^\alpha},a^\alpha\in\A^\alpha}$ map a state $x$ and action $a^\alpha$ to every possible next state $x'$, taking into account the probability of the teammate executing each action on $x$. The reward function $r$ assigns the reward of $100$ to states $x$ where the cook delivers a cooked soup through the kitchen window, and $-1$ otherwise. After the soup is delivered, the task is considered complete, leading to an absorbent state which is assigned a reward of $0$. We consider a discount factor $\gamma=0.95$.

\clearpage

\begin{table*}
    \centering
    \caption{Hyperparameters used for the Perseus \citep{spaan2005perseus} algorithm. The discount factor was also used for the Value Iteration algorithm.}
    \begin{tabular}{lcccc}
    
        \toprule
        
        \bf Environment & \bf Horizon & \bf Beliefs & \bf Tolerance & \bf Discount \\
        \bf             & \bf         & \bf         & \bf           & \bf Factor \\
        
        \midrule
        
        gridworld & 50 & 5000 & 0.01 & 0.95 \\
        pursuit-task & 75 & 5000 & 0.01 & 0.95 \\
        pursuit-teammate & 85 & 5000 & 0.01 & 0.95 \\
        pursuit-both & 85 & 5000 & 0.01 & 0.95 \\
        abandoned power plant & 50 & 2500 & 0.01 & 0.95 \\
        ntu & 75 & 5000 & 0.01 & 0.95 \\
        overcooked & 50 & 1800 & 0.01 & 0.95 \\
        isr & 75 & 5000 & 0.01 & 0.95 \\
        mit & 75 & 5000 & 0.01 & 0.95 \\
        pentagon & 75 & 5000 & 0.01 & 0.95 \\
        cit & 85 & 8000 & 0.01 & 0.95 \\ 
        
        \bottomrule
        
    \end{tabular}
    \label{table:hyperparameters}
\end{table*}

\begin{figure*}[!tb]
    \centering
    \includegraphics[width=.65\linewidth]{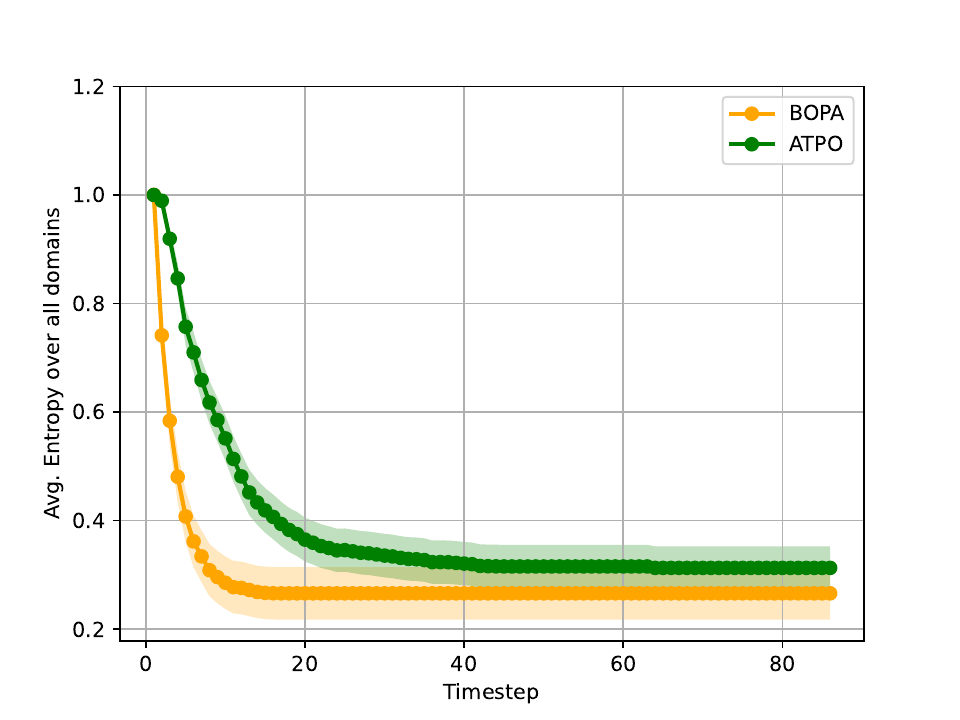}
    \caption{Average belief entropy over all environments for ATPO and BOPA. Both Bayesian inference agents are successfully able to identify the correct model as they interact with the environment. Compared to BOPA, ATPO, on average, requires extra timesteps to identify the correct model. This is an expected result, given that BOPA is able to fully observe the environment and therefore has less uncertainty when performing the inference.}
    \label{fig:entropy}
\end{figure*}

\end{document}